\documentclass[12pt]{article}
\usepackage{vmargin}
\usepackage{pa}
\usepackage{amsmath}
\usepackage{amsxtra}
\usepackage{amsbsy}
\usepackage{amssymb}
\usepackage{indentfirst}
\usepackage{array}
\usepackage{dcolumn}
\usepackage{bm}
\usepackage[longnamesfirst,sort]{natbib}
\usepackage{booktabs}
\usepackage[pdftex]{graphicx}
\usepackage{xcolor,colortbl}
\definecolor{lg}{gray}{.8}
\definecolor{dg}{gray}{0.6}
\usepackage{subcaption}
\usepackage[pdftex=true,%
            pdftoolbar=true,%
            pdfmenubar=true,%
            pdfauthor = {Nathaniel\ Beck},%
            pdfcreator = {PDFLaTeX},%
            pdftitle = {Binary DVs},%
            colorlinks=true,%
            urlcolor=blue,%
            linkcolor=blue,%
            citecolor=blue,%
            implicit=true,%
            hypertexnames=false%
            ]{hyperref}
 \usepackage{pdfsync}
\usepackage{url}
\usepackage{xspace}
\usepackage{numprint}
\usepackage{longtable}
\newcommand{\newc}{\newcommand}
\newc{\skipc}[1]{\multicolumn{#1}{c}{}}
\newc{\N}{\hphantom{0}}
\newc{\M}{\hphantom{$-$}}
\newc{\barr}{\begin{array}}
\newc{\earr}{\end{array}}
\newc{\bcen}{\begin{center}}
\newc{\ecen}{\end{center}}
\newc{\subsect}{\subsection*}
\newc{\beqann}{\begin{eqnarray*}}
\newc{\eeqann}{\end{eqnarray*}}
\newc{\beqa}{\begin{eqnarray}}
\newc{\eeqa}{\end{eqnarray}}
\newc{\beqnn}{\begin{displaymath}}
\newc{\eeqnn}{\end{displaymath}}
\newc{\beq}{\begin{equation}}
\newc{\eeq}{\end{equation}}
\newc{\tabst}{\begin{table}\centering}
\newc{\tabstp}{\vfill\begin{table}[p]\centering}
\newc{\tabend}{\end{tabular}\end{table}}
\newc{\btab}{\begin{tabular}}
\newc{\fn}{\footnote}
\newc{\cols}{\multicolumn}
\newc{\vm[1]}{\bm{\mathrm{#1}}}
\newc{\ea}{et al.}

\newcommand{\bb}{\ensuremath{\pmb{\beta}}\xspace}

\newcommand{\gb}{\ensuremath{\beta}\xspace}

\newcommand{\bX}{\ensuremath{\pmb{X}}\xspace}

\newcommand{\bbh}{\ensuremath{\hat{\pmb{\beta}}}\xspace}
\newc{\tsl}{\textsl}
\newc{\bit}{\begin{itemize}}
\newc{\eit}{\end{itemize}}
\newcommand{\bxgi}{\ensuremath{\pmb{x}_{g,i}\xspace}}
\newcommand{\ygi}{\ensuremath{y_{g,i}\xspace}}
\title{Estimating  grouped data  models with a binary dependent
  variable and fixed effects: What are the issues?}
\author{Nathaniel Beck\thanks{Department of Politics;
           New York University;  New York, NY 10003 USA;
           \texttt{nathaniel.beck@nyu.edu}. Simulation code is
           available in \citen{nb:spm15dvn}.
          For the few replication
         examples, please contact the original authors for their
         data.} 
       \date{\today}}

\date{\today.}

 \newcommand{\citen}{\citet}
 \renewcommand{\cite}{\citep}

\newcolumntype{.}{D{.}{.}{-1}}
\newcolumntype{d}[1]{D{.}{.}{#1}}
\newcolumntype{E}{D{.}{.}{3}}

\setpapersize{USletter}
\setmarginsrb{1in}{1in}{1in}{1in}{0pt}{0mm}{0pt}{0.25in}

\begin{document}

\maketitle
\pagebreak

\begin{abstract}
This article deals with a very simple issue: if we have
grouped data with a binary dependent variable and want to include
fixed effects (group specific intercepts)  in
the specification, is Ordinary Least Squares (OLS)  in any way
superior to a (conditional) logit form? In
particular, what are the consequences of using OLS instead of a fixed
effects logit model with respect to the latter dropping all units which
show no variability in the dependent variable while the former allows
for estimation using all units. First, we show that the discussion of
fixed effects logit (and the incidental parameters problem) is based
on an assumption about the kinds of data being studied; for what
appears to be the common use of fixed effect models in political
science the incidental parameters issue is illusory. Turning to linear
models, 
we  see that OLS yields a perhaps odd
linear combination of the estimates for the units with and without variation in
the dependent variable, and so the
coefficient estimates must be  carefully interpreted. The
article then compares two methods of estimating logit models with
fixed effects, and shows that the Chamberlain conditional logit is as
good as or better than a logit analysis which simply includes group
specific intercepts (even though the conditional logit technique was
designed to deal with the incidental parameters problem!).  Related to
this, the article discusses the estimation of marginal effects using
both OLS and logit. While it appears that a form of logit with fixed
effects can be used to estimate marginal effects, this method can be
improved by starting with conditional logit and then using the
those parameter estimates   to constrain the logit with fixed
effects model. This method produces estimates of sample average
marginal effects that are at least as good as OLS, and much better
when group size is small or the number of groups is large. 
These issues are simple to
understand, but it appears that applied researchers have not always
taken note of them.
\end{abstract}
\newpage
\section{Introduction}
\label{sec:intro}
Many applied researchers include ``fixed effects'' (unit specific intercepts) to account for unmodeled
heterogeneity in grouped data  analyses; these fixed effects lead to
interesting issues.\fn{Researchers using data with a temporal
  component often also include temporal effects. These typically do
  not cause the problems of group fixed effects, since the number of
  temporal effects is usually small and the number of observations per
  temporal effect is usually quite large. Thus fixed effects in this
  article are only  group specific intercepts and not time specific
  intercepts.}
This is a well worked area when the dependent
variable is continuous (see, any standard econometrics text, such as
\citeauthor{cameron_trivedi:bk}, \citeyear{cameron_trivedi:bk},
ch. 21, or \citeauthor{greene:bk}, \citeyear{greene:bk}, ch. 9).  The situation is more
complicated when the dependent variable is binary, though again the
theory is well worked out (see \citeauthor{cameron_trivedi:bk}, \citeyear{cameron_trivedi:bk},
ch. 23 or   \citeauthor{greene:bk}, \citeyear{greene:bk}, ch. 23). In
particular, the group mean centering solution for 
estimating a model with fixed effects and a continuous
dependent variable does not carry over to non-linear models, such as
logit.\fn{Most work in this area uses logit and at least some results,
  such as those of \citen{chamberlain:restud}, do not carry over to
  probit and so this article only considers the logit specification;
  many results carry over to the probit specification but this is of
  little interest. To keep specification and estimation distinct I
  refer to the generic logit specification with fixed effects as ``LOGITFE'' regardless of how it is estimated.}

 It is, of course, possible to estimate a logit
specification with group specific intercepts (dummy variables); this
method (denoted as ``FELOGIT'' to keep method and specification distinct) has not been heavily used, partly for computational reasons and
partly because of a misunderstanding about relevant asymptotics. This
misunderstanding is treated in Section~\ref{s:neyman}. Researchers
working with the LOGITFE specification have instead turned to
\citeauthor{chamberlain:restud}'s \citeyearpar{chamberlain:restud}
conditional logit (denoted ``CLOGIT'' to again keep method and
specification distinct), which does provide consistent estimates under
some, perhaps irrelevant for a given researcher,
conditions. Section~\ref{s:neyman} makes it clear which asymptotics
are relevant. 
 
As is well known, either logit approach 
 implies that groups with no variation on the dependent variable
contain no information that help identify the parameters, and these 
observation do not enter the likelihood function. Alternatively many
researchers resort to the
the simpler linear probability model (hereinafter ``LPMFE'') estimated
by OLS\@, 
which appears to use observations from all  groups 
to estimate parameters. This article begins by unpacking 
 the relationship between
the LPMFE and LOGITFE models with respect to this  change in the data
set used for estimation. Section~\ref{s:diff} deals with this issue. Social
science data sets often contain  many groups with no variation on the
dependent variable, 
 This change in the data being analyzed is
often unremarked upon, but obviously in some research 
contexts can be consequential. 

Another reason that researchers often turn to LPMFE is that it allows
for sample average marginal effects, which require the estimation of
the various effects. Since CLOGIT just conditions out those effects,
it cannot provide estimates of sample marginal effects. FELOGIT can
provide such estimates, though many researchers appear to have been
reluctant to use it because of the misunderstood asymptotic
issues. Various researchers \cite{katz:pa, greene:fe,coupe:pa} have shown that
the bias in using FELOGIT to estimate substantive parameters of
interest is small in practical situations, that is
when group size is at least twenty. However, on the way to discussing
the estimation of sample marginal effects, this article reexamines the
issue of the accuracy of FELOGIT and GLOGIT in terms of mean squared
error and not bias; in Section~\ref{s:accuracy} it is shown that
CLOGIT is more accurate than FELOGIT even in situations where FELOGIT
is essentially unbiased.

  Section~\ref{s:marginal} returns  to   the estimation of
 sample average marginal effects. Is 
FELOGIT sufficiently accurate so that the estimated sample marginal
effects are meaningful, or more correctly, how large do group sizes
have to be before such estimates are meaningful.This section discussed
an improvement on FELOGIT which builds on the previous section's
comparison of FELOGIT AND CLOGIT and then compares this improved
estimator to the OLS estimates of sample marginal effects implied by
the LPMFE, showing that, at least in one case, the improved estimator
of the LOGITFE specification is superior to OLS estimation of the
LPMFE specification. 

Since this article only  deals with data where the group size is large
enough to make the estimation of fixed effects at least plausible, is
this argument relevant to political science as practiced? Similarly,
are researches mixing the LOGITFE model with the LPMFE model without
sufficiently considering the consequences? The answer to both
question is yes. Large but not huge  group sizes (20-100)  are common
in political science; the issue of
fixed effects is also common in such models. While much research
involves continuous dependent variables (for which this article is
irrelevant), there is a non-trivial amount of research where the
dependent variable is dichotomous. Evidence of this is provided by a
search of JSTOR.
 
From 2000-2015 (June), a search on
``linear probability'' and ``fixed effects'' found 1158 articles, of
which 87 were in political science or international relations (the
majority, 798 articles, were in economics). When ``conditional logit''
was added to the search term, the total number of
articles fell to 86 (with only 7 in political science or international
relations). Many research articles, for whatever reason, fit a linear
probability model when the specification includes fixed
effects. Interestingly, this is about half of all articles that use the term
``linear probability;'' there were 2180 articles returned with just this one
search term; for political science and international relations the
corresponding figure is about one third of the 276 articles which used the
term ``linear probability.''\fn{For completeness, a similar search
  found 449 articles with the search terms ``fixed effects'' and
  ``conditional logit,'' with 49 of those being in political science
  or international relations. But researchers refer to both the
  Chamberlain procedure and McFadden's conditional (multinomial) logit; 147 of the
  449 articled cited McFadden (but not Chamberlain) whereas 100 cited
  Chamberlain but not McFadden. Alas, that leaves 202 articles citing
  neither. Political scientists were even less likely to cite either
  McFadden or Chamberlain, with the 16 who did cite one or the other
  evenly split between them. Based on this, it does appear that
  researchers fitting a model with a binary dependent variable and
  fixed effects gravitate to the linear probability model rather than
  the Chamberlain conditional logit model.} Finally, at least amongst
the political science articles, I did not find articles which
estimated the LPMFE specification on large behavioral panels, that is,
ones with many respondents (groups) and very few waves (group
size). Such may exist, but they are at best uncommon in political
science.

Many authors do not clarify why they chose to use the linear
probability model. One reason is for simplicity in dealing with
endogenous regressors where the linear probability model is much
simpler to estimate \cite{angrist:jbes} ; 836 articles using the linear probability model
also used the term endogeneity; somewhat more than half (495) also
included fixed effects; in political science and international
relations the corresponding article counts are 73 and 38. This leaves
many articles which estimate linear probability models (both with and
without fixed effects) where there are no endogeneity issues.

I had hoped that a portion of articles which estimated linear
probability models with fixed effects would also have mentioned issues
related to either not having to drop group due to lack of variability
in the dependent variable or because of a desire to estimate marginal
effects. But having read some of the more prominent articles  in major
journals, I find that authors tend to either report both LMPFE and CLOGIT
results, simply remarking that the results are not much different, or
make the same remark but put the second set of  results in an group of
``robustness checks'' or  report only one set of  results. 

To get more detail,  take the three most recent pieces using the
LPMFE model in the \emph{American Political Science
  Review}. \citen{hainmueller:apsr} provides an estimate of the
probability of an application for naturalization in Switzerland being
rejected as a function of individual characteristics and a
municipality 
fixed effect; there are about fifty applications per municipality. Their
specification is the LPMFE with no discussion of non-linear
alternatives.\fn{In fairness, this analysis  is secondary to an
analysis of the proportion voting to reject the naturalization
petition. However, they fit a linear model to the proportion voting to
reject which is only consistent with a linear probability model.}
 \citen{besley:apsr} studies whether democracies provides
more educated leaders by estimating models, for example, of the
probability of a leader having a graduate degree in a large number of
countries, where the specification includes country fixed
effects. This article provides both OLS estimates of the LPMFE
specification and CLOGIT estimates of the LOGITFE
specification. \citen[559]{besley:apsr}
only mentions the non-linear estimate by noting that ``[In the LOGITFE
specification], we estimate a conditional logit
model to recognize the discrete nature of the lefthand-side
variable. The core finding of [the LPMFE model] remains.''  This is
clearly correct if we only care about the sign and significance of a
coefficient, but, as we shall see, the difference between the two
estimations is not trivial albeit not enormous.  Finally,
\citen{petrova:apsr} estimates a LPMFE specification 
for  newspaper independence as a function of
profitability over a half decade in the 1880's, sometimes grouping the
observations by newspaper (so about 5 observations per group) and
sometimes by county (with a much larger number of observations per
group); this article is relevant to the latter specification. As in the
\citen{besley:apsr} piece, the only methodological discussion of this
is in a footnote (p. 796) which states ``[t]he results of estimating
the fixed-effect conditional logit are consistent with the results of
fixed-effects OLS, as discussed in the robustness check
section.''\fn{This discussion is then well hidden.}
 In short, real articles in top journals  use methods discussed here,
 and use them without much methodological
justification. This article seeks to provide some methodological
clarity. It begins by laying out the notation used.

\section{Notation}

Let  \ygi\ be a binary dependent variable with the exogenous
covariates being
\bxgi, where 
 $g$ indexes groups and $i$ indexes particular units in a group.
  It simplifies notation to assume that all groups are of the
same size, and dropping this one extra subscript has no consequences
for the argument:  let this be group size be $N$, with $G$  being  the
number of groups. Let the number of covariates be $k$; in this article
everything holds even when $k=1$ and it is does not matter whether we
think of the covariates as a  vector or a scaler. $f_g$ refers to the fixed effect for group
$g$, that is the group specific intercept.

 What is critical for this article is that $G$ is
fixed; 
asymptotics are in terms of $N$. In the articles cited in the previous
section, neither the number of municipalities, the number of countries nor
the number of counties can be thought of as going to infinity; the
number of groups may be large (counties), but
the critical thing is that the number of groups is fixed, so that the
asymptotic properties of estimators considered here are in terms of
$N$, not $G$. 

The data may take any grouped form, such as 
time-series--cross-section or simply observations grouped by some unit
(village, tribe, state, country); even where the data has a structure
so that observation $i$ means the same thing in different groups, this
structure is orthogonal to the arguments of this article (though obviously researchers would
need to take account of that structure in their analyses). The article distinguishes groups that
have variation on the dependent variable from those that do not. For
convenience it is assumed that groups with no variation (``ALL0'') have
$\ygi=0$.\fn{It is trivial to extend the argument to data sets which
contain some all failure and some all success groups. In
the LPMFE specification the estimates for the all successes groups are
identical to those for the all failures group except for a constant
term. 0 is used interchangeably with failure and 1 with success.} 
these are denoted ``ALL0'' groups.  Assuming
  exogeneity of the covariates, the generic model with fixed effects
  is 
\begin{equation}
P(\ygi=1) = H(\bxgi\bb + f_{g'}I_{g'=g})
\end{equation}
where H is some (possibly stochastic) function which needs to be specified; different H's
lead to different specifications. I is the usual indicator function
which in this case indicates group membership. 

The LMPFE is obtained by setting  H so that 
\begin{gather}
\ygi = \bxgi\bb + f_{g'}I_{g'=g}+ \epsilon_{g,i} \label{eq:linear} \\
P(\ygi=1) = E(\bxgi\bb + f_{g'}I_{g'=g}) + \epsilon_{g,i}) = \bxgi\bbh +
\widehat{f_i}I_{g=k}.  \nonumber
\end{gather}
This can be estimated by OLS and it is assumed that $\epsilon_{g,i}$  satisfies the Gauss-Markov assumptions. Of course these
    ``probabilities'' need not be between zero and one, and this
    specification suffers from all the standard issues related to the
    linear probability model in general. 
The  LOGITFE specification  is obtained by choosing H so that 
\begin{equation}
Pr(\ygi=1)=\frac{1}{1+e^{-(\bxgi\bb + f_{g'}I_{g'=g})}} 
 \label{eq:logit}
\end{equation}
where the use of $\bxgi\bb$ indicates that the covariates combine
to affect $y$ through a single-index model
\cite[123]{cameron_trivedi:bk}. The estimation of this model has been
the subject of much theoretical discussion sparked by
\citeauthor{neyman:econ}'s \citeyearpar{neyman:econ} discussion of the
``incidental parameters problem'' almost 70 years ago.  

\section{The incidental parameters ``problem''} \label{s:neyman}

In both specifications the number of parameters is $G+k$, whereas the
number of observations if $NG$. Now while it may not be advisable to
estimate any model where the number of parameters is a sizable
fraction of the number of observations (which happens when $N$ is
small), there is no violation of any standard assumptions in this
situation. Of course parameter estimates may be inaccurate for small
$N$, and we know that logit models do not produce unbiased estimators
in finite samples. As we shall see, estimating logit models with a large
number of parameters (relative to the number of units) is problematic,
but this issue is orthogonal to the issue that $G$ of the parameters are
group specific intercepts. 

There are situations where the asymptotics are in $G$, not
$N$. This would be the case for ``behavioral panel'' data, where,
there are  a few, and fixed, number of interview
waves and asymptotics are in terms of the number of people
interviewed.  It is hard in political science to find such
studies when the dependent variable is binary; if one has such data,
the conclusions of this article are irrelevant. The
distinction between the type of data considered here (fixed $G$) and
the behavioral panel data has, however, led to confusion.

In particular, if the number of intercepts to be estimated goes to
infinity as the number of observations goes to infinity, as in the behavioral panel case, we get  the
``incidental parameters'' problem; in this
situation standard maximum likelihood results do not
hold and maximum likelihood estimators may not even be consistent.  \citeauthor{neyman:econ} showed, however, that if one could
estimate parameters of interest conditional on the incidental
parameters that the parameters of interest would then be consistently
estimated. For the fixed effects with a continuous dependent variable
situation and a linear specification, this conditioning consists of group mean centering all
observations. The standard
texts mentioned in the introduction  easily show that, because of the linearity of the specification, estimating
the linear model with the fixed effects using  OLS is identical to this
conditional estimation. Alas, this result depends heavily on linearity
and when we move to the non-linear world things get more
complicated. In particular, standard logit estimation of the LOGITFE
specification are inconsistent \emph{for behavioral panel data with
  asymptotics in $G$}.

For the types of data considered here, there is literally no
incidental parameters problem, and one could simply estimate the
LOGITFE specification by standard logit (FELOGIT). While the LOGITFE
specification may contain a large number of parameters, this number
does not grow with $N$. This is, again, not
to say that the large number of parameters to be estimated is not
problematic, but the issue is the large number of parameters relative to
the number of observations in non-linear models and unrelated to the
fact that these parameters are group specific intercepts. 

\citen{chamberlain:restud} showed that $\bb$ can be
consistently estimated for the behavioral panel case (where incidental parameters
are  an issue) by conditioning on the number of successes in
any group, that is, given a group with k successes, estimate
$\bb$ by finding the value that best predicts which  units in that
group are successes (CLOGIT). There is nothing wrong with doing this for
the types of data considered here, but there is no need to do so in
terms of asymptotic theory.\fn{An excellent discussion of this issue, including
  various mis-applications of the Neyman-Scott argument may be found 
  in \citen{greene:fe} written over a decade ago. This article, which is surely under-cited in
  political science, points out, as pointed out here, that finite and
  non-growing $G$ makes the Neyman-Scott problem irrelevant. The
  article also points out that some of the received wisdom on fixed
  effects and non-linear models overstates the problem. The issues
  considered in this article are somewhat different from those considered by
  Greene, though the finite sample queries are similar in spirit.}

Because of the non-linearity of the logit specification, 
 CLOGIT  and FELOGIT are not identical. Note that
CLOGIT conditions on a known number, the number of ones in a group;
FELOGIT estimates all the fixed effects, and the imprecision of those
estimates  ``leaks''  into the estimation of the parameters of
interest. As we shall see, CLOGIT may well outperform FELOGIT, but
this 
is because of the imprecisely estimated parameters in the FELOGIT
model, not the Neyman-Scott issue. This is dealt with in
Section~\ref{s:accuracy}. The article now begins with a comparison of
LOGITFE and LPMFE in terms of which observations are dropped in the
estimation of \bb.

\section{Differences between what is estimated with LPMFE and
  LOGITFE} \label{s:diff}

As noted, any of the methods used to estimate a LOGITFE specification
drop all the ALL0 groups (or, alternatively, these groups do not enter
into the likelihood). The LPMFE model estimated by OLS does use
information on all the groups. To see the consequences of this, note
that the OLS estimate of \bb is a weighted average of the estimates in
the ALL0 and the other (``NOTALL0'')  groups.\fn{Any estimator can be
  seen as a combination of estimators for subgroups of data; this is a
  long standing idea in econometrics, with perhaps the best known and
  long standing example being the the Chow test of the equality of
  regression lines in two subsets of data \cite{chow:econ}. The
  calculations here are even simpler because of the nature of the
  dependent variable in the ALL0 group.}
 For the ALL0 groups,
$\ygi=0$ so the OLS estimate of \bb is zero.\fn{All group intercepts in
the ALL0 group are also zero, and the fit appears to be
perfect. This of course follows simply because the group mean centered
\ygi's in the ALL0 groups are also always zero.
It is also necessary to assume  some variation in the
covariates for the ALL0 groups so that $\pmb{\tilde{\mathbf{X}}'\tilde{\mathbf{X}}}$  is
not singular for the ALL0
groups.} Let $\pmb{\tilde{\mathbf{X}}}$ and $\pmb{\tilde{\mathbf{y}}}$ be the group mean centered
analogues of $\pmb{\mathbf{X}}$ and $\pmb{\mathbf{y}}$ so we can avoid worrying
about the fixed effects in the OLS estimations.
Let $\pmb{\tilde{\mathbf{X}}_0}$ be the covariate matrix for the ALL0
groups with $\pmb{\tilde{\mathbf{X}}}_1$ being the corresponding matrix for the
NOTALL0 groups.  Similarly,  $\pmb{\tilde{\mathbf{y}}}_1$ is  the
group mean centered  vector of
observations on $y$ for the NOTALL0 group with the corresponding vector
for the ALL0 group being $\pmb{\mathbf{0}}$. Thus the OLS estimate of \bb
for the entire data set ($\bb_{01}$)  are given by 
\begin{equation}
\bbh_{01} =
\pmb{(\tilde{\mathbf{X}}_1'\tilde{\mathbf{X}}_1+\tilde{\mathbf{X}_0}'\tilde{\mathbf{X}}_0)^{-1}(\tilde{\mathbf{X}}_1'
\tilde{\mathbf{y}}_1)}
\label{eq:b01}
\end{equation}
whereas the corresponding estimate for the NOTALL0 
groups ($\bb_1$)
is given by 
\begin{equation}
\bbh_1=
\pmb{(\tilde{\mathbf{X}}_1'\tilde{\mathbf{X}}_1)^{-1}(\tilde{\mathbf{X}}_1'
\tilde{\mathbf{y}}_1)}.
\label{eq:b1}
\end{equation} 
We can also compare the variance covariance matrix of the two
estimates. For the entire data set this matrix is  
\begin{equation}
\pmb{(\tilde{\mathbf{X}}_1'\tilde{\mathbf{X}}_1+\tilde{\mathbf{X}}_0'\tilde{\mathbf{X}}_0)^{-1}}\widehat{\sigma_{01}^2}
\label{eq:s01}
\end{equation}
whereas the corresponding estimate for the NOTALL0 
groups ($\bb_1$)
  is given by 
\begin{equation}
\pmb{(\tilde{\mathbf{X}}_1'\tilde{\mathbf{X}}_1)^{-1}}\widehat{\sigma_{1}^2}
\label{eq:s1}
\end{equation}
where $\widehat{\sigma_{01}^2}$ and $\widehat{\sigma_{1}^2}$ refer to
estimates of the standard error of the regression in the full and
restricted data sets respectively.

It is immediately obvious that the two equations only differ by the 
$\pmb{\tilde{\mathbf{X}}_0'\tilde{\mathbf{X}}}_0$ portion of the $\bX'\bX$ matrix that is being
inverted. Alternatively, it is obvious that the OLS estimates for all
the data is a weighted average of $\mathbf{0}$ and the $\bb_1$;  $\bb_{01}$ shrinks
$\bb_1$ towards $\mathbf{0}$.  The amount of shrinkage is a 
 somewhat complicated function that depends on the
relative scale of $\pmb{\tilde{\mathbf{X}}_0'\tilde{\mathbf{X}}_0}$ and
$\pmb{\tilde{\mathbf{X}}_1'\tilde{\mathbf{X}}_1}$. The difference between $\bbh_1$ and
$\bbh_{01}$ is similar.\fn{I use the term scale here because
  there is no simple measure of the ``size'' of
  $\pmb{\tilde{\mathbf{X}}_0'\tilde{\mathbf{X}}_0}$ in general.}

The variance covariance matrix of the estimates has two components
which move in different directions as we move from the entire data set
to the NOTALL0 data set. The estimated $\sigma^2$ will get smaller,
since we are eliminating non-homogenous cases; however 
the $\pmb{\tilde{\mathbf{X}}_1'\mathbf{\tilde{X}}_1}$ matrix in the NOTALL0 data
 will also be smaller
in scale than the corresponding
 $\pmb{\tilde{\mathbf{X}}_{01}'\mathbf{\tilde{X}}_{01}}$ 
matrix used to estimate
the variance covariance matrix of of $\bbh_{01}$. Note however that the estimated standard error of the
regression will be limited in how much it changes since the variance
of $\pmb{\tilde{\mathbf{y}}}$ is limited by it being a binary variable; the $\pmb{\tilde{\mathbf{X}}'\tilde{\mathbf{X}}}$ matrix is
not similarly limited by any scaling, and so could shrink considerably
as the ALL0 cases are dropped (depending of course on how many ALL0
groups there are and the scale of the $\pmb{\tilde{\mathbf{X}}'\tilde{\mathbf{X}}}$ matrix for those groups.).
 In general, the estimated standard errors of $\bbh_{1}$ will be
 smaller than the corresponding estimates for $\bbh_{01}$ The
change in \bbh\  and its estimated standard error
offset (in general), and so we usually see smaller impacts on the $t$-ratio
associated with $\bb_{01}$ as compared to $\bb_1$, even though both
components of the ratio may change more markedly; this may be one reason
that authors are content to conclude that the substantive results from
LOGITFE  are similar to those of LPMFE. But we should go beyond simply
inquiring as to the sign of a coefficient and whether its
``significance'' is beyond some standard threshold.

It is very simple to see what is going on by looking at the scalar $x$
case, where once again $\tilde{y}$ and $\tilde{x}$ have been group mean centered. The OLS estimate of $\gb_{01}$ 
for the entire data set is given by 
\begin{equation}
\hat{\beta}_{01} =
\frac{\sum\limits_{\text{NOTALL0}} \tilde{x}_{g,i}\tilde{y}_{g,i}}{\sum\limits_{\text{ALLDATA}} \tilde{x}_{g,i}^2} 
\label{eq:sb01}
\end{equation}
whereas the corresponding estimate for the NOTALL0 
groups ($\gb_1$)
  is given by 
\begin{equation}
\hat{\beta}_{1} =
\frac{\sum\limits_{\text{NOTALL0}} \tilde{x}_{g,i}\tilde{y}_{g,i}}
{\sum\limits_{\text{NOTALL0}} \tilde{x}_{g,i}^2}.
\label{eq:sb1}
\end{equation}
These  two equations  differ only  by an extra $\sum\limits_{\text{ALL0}}
\tilde{x}_{g,i}^2$ in the denominator of Equation~\ref{eq:sb01}; this extra
term so $\hat{\gb}_{01} < \hat{\gb}_1$. 
The standard error for $\hat{gb}_{01}$ for the entire data set is  given by
\begin{equation}
\sqrt{\frac{\widehat{\sigma_{01}^2}}{\sum\limits_{\text{All Data}} \tilde{x}_{g,i}^2}}
\label{eq:ss01}
\end{equation}
whereas the corresponding standard error  for the NOTALL0 
groups ($\hat{\gb_1}$)
  is given by 
\begin{equation}
\sqrt{\frac{\widehat{\sigma_{1}^2}}{\sum\limits_{\text{NOTALL0}} \tilde{x}_{g,i}^2}}
\label{eq:ss1}
\end{equation}
where again the extra summation terms in the denominator must be positive.

For the scalar case it is obvious that including the ALLO groups 
shrinks $\hat{\gb_1}$ towards zero, where the amount of shrinkage
depends on how many ALL0 groups there are and the variation of the
centered $x$'s in those groups. The estimated standard error of $\gb_1$
also gets smaller (in general), since the larger denominator due to
$\sum\limits_{\text{ALL0}} \tilde{x}_{g,i}^2$ will almost always offset the increase
in the estimate of the standard error of the regression due to the
greater heterogeneity of y of the full data set. This again leads to
offsetting effects in changing $t$-ratios.

To see how this works in practice, we can look at the regression
results of both \citen{besley:apsr} and
\citen{hainmueller:apsr}.\fn{The published results in both articles were
trivial to replicate. Here I focus on one important variable in each
study.} The \citeauthor{besley:apsr} estimate for the effect of democracy
on whether a leader had a graduate degrees (Table 1, Column 1) 
 was $0.22$ with a standard
error of $.048$ using all 1146 observations. When limiting the analysis
to the 956 observations in the NOTALL0 groups, the corresponding
estimates are $0.26$ and $0.051$. 

The corresponding change for
\citeauthor{hainmueller:apsr} is similar. Using the results of 2429
applications for naturalization in Switzerland, their estimate for the
effect of being from the former Yugoslavia on rejection of a
naturalization claim (Table 3, Column 2) was about $0.30$ with a
standard error of $.05$; this figure rises to about $0.36$ with a
standard error of $.06$ when the 408 ALL0 municipalities are dropped. 

In both cases marginal effects including ALL0
countries understates marginal effects by about 15--20\% with a change in
$t$-ratio of under 10\%. While the effect is far from enormous, such
changes are not trivial when we consider the complicated methods we
use to get small increases in efficiency of estimation. And, of
course, the effects may be much larger if the number of ALL0 groups is
larger than in these two cases.\fn{The largest number of cases dropped
 due  to ALL0 groups in published analysis is the \citen{green:io} fixed effects analysis of
  Militarized Interstate Disputes, where 93\% of the data does not
  enter the likelihood function. Whether it makes any sense to include
  fixed effects for data like this is discussed in \citen{nbkatz:io}
  and this issue is not discussed further here.}

To correctly compare LOGITFE results with LPMFE results, clearly the
latter specification should be estimated dropping the ALL0 groups. But
one can make a case that the the LPMFE results using all groups, which
average zero with the LPMFE on the restricted data set, make sense in
that the marginal effect of the covariates on $y$ could be thought of
as being zero in the ALL0 groups. After all, if $P((\ygi=1)=0|g\in
ALL0$), then then marginal effect of any $x$ in the ALL0 groups is
indeed zero. Alternatively, we can think of this as a meaningless
exercise, since some change in an $x$ in an ALL0 group member will
change a failure to a success and thus its marginal effect cannot be
zero. Researchers can report both numbers and their interpretation;
what is clear is that researchers must understand the difference
between the two estimates, and understand how to compare LOGITFE and
LPMFE results.\fn{It also might look like the issue is related to a
  tobit type model, where the ALL0 groups look like the tobit
  zeros. One might think of it this way, but the presence of fixed
  effects, which perfectly explain the ALL0 groups, makes analysis
  along these lines impossible. Unlike the tobit case, the fixed
  effects leave nothing in the error term to be integrated over,
  whether the underlying model is linear or logit.}

\section{CLOGIT VS FELOGIT} \label{s:accuracy}

As noted in the previous section, researchers have typically estimated the
LOGITFE specification using CLOGIT out of fear that FELOGIT is
inconsistent. To repeat
what was shown there,  if $G$ is fixed, and asymptotics
are in $N$, there is literally no incidental parameters problem, and
FELOGIT is consistent (as $N \rightarrow\infty$), as of course is CLOGIT.Note that, unlike the
continuous $y$ case, conditioning on the fixed effects is not identical
to including them in the specification. Both methods drop ALL0 groups, but FELOGIT does
allow for estimating sample marginal effects (which required
estimation of the fixed effects). But before looking at FELOGIT
estimates of the marginal effects, it is necessary to compare the
finite sample properties of FELOGIT vs. CLOGIT for estimating the
parameters of interest, \bb. CLOGIT may well outperform FELOGIT
because it conditions on the known number of successes in a group,
rather than (imprecisely) estimating the fixed effect for that
group. As $N \rightarrow \infty$ the two estimators must converge
(they are both consistent in $N$), but how do the two estimators
compare in finite samples. 
To answer this we must turn to Monte Carlo simulations. These simulations
will consider various values for $G$ in situations observed in actual
research (mid to high two figures) and vary $N$. 

\citen{katz:pa} and \citen{coupe:pa} show that the bias in FELOGIT is 
small when $N>16$ though when $N$ is small the bias is large (100\%
when $N=2$); CLOGIT is essentially unbiased in all their reported
results.  However, even though FELOGIT may become unbiased for
relatively small $N$, this does not mean it is as accurate as CLOGIT
for such $N$. What is important is not unbiasedness, that is, whether
the average of the \bbh\ over the simulations is close to the
known \bb, but rather accuracy, that is, how close are each of the
\bbh\ to the known \bb.\fn{Those with a touching faith in unbiasedness
  can skip this section, since the various simulation studies cited
  above tell us that bias is not an issue for FELOGIT for group size
  much above 20, and that CLOGIT is essentially always
  unbiased. And for those, like me, who find unbiased uninteresting,
  it must be remembered that RMS error is the sum of squared bias plus
  variation centering on the estimated parameter, so bias contributes,
  sometimes non-trivially, to inaccuracy. 
Also note that I use the term accuracy rather than
  efficiency since the latter often  refers only to the class of
  unbiased estimators.}  The simulations have only a single parameter of
interest, \gb (nothing changes if there are a few parameters of
interest). 
To compare the accuracy of FELOGIT and CLOGIT,
we   look at the ratio of the root mean squared  errors (RMSEs) of \bbh\
from FELOGIT and 
CLOGIT. To be precise, the RMSE of a scalar  estimator (in $R$ simulation
runs) is
\begin{equation}
\sqrt{\frac{\sum_{r=1}^R (\hat{\gb}_r-\beta)^2}{R}}.
\end{equation}

Data were simulated according to Eq.~\ref{eq:logit} using a scalar
$x$. $G$ was taken as either 20, 50 or 100 
 and $N$  was varied between 3 and 100 as in
 Table~\ref{tab:ineff}. Fixed effects
were drawn from a standard normal distribution;  $x$ was generated to be
correlated with the fixed effects by adding together a standard normal
and some fraction of the group fixed effects (yielding a $R^2$ of the
regression of $x$ on the fixed effects of about $.20$); 
$\beta$ was set to one and there was no overall constant term (so the
expected value of the latent for $y$ was zero); after probabilities were generated according
to Eq.~\ref{eq:logit}, a realized value of 0 or 1 was drawn for each
observation using the Bernoulli distribution. Given these parameter
values, the average probability of success was 0.5 with the individual 
probabilities of success  distributed symmetrically around this value.\fn{All simulations were
  done using Stata 14. For the computationally interested, both the
  CLOGITs and FELOGITs take between a tenth and half a second each on
  a well equipped iMac. All simulation results were based on 1000
  simulations; in cases where one of the maximum likelihood routines
  did not converge, another data set was drawn so all results average
  1000 analyses. Different parameter values were tried but all yielded
qualitatively similar results and so these numerous simulation results are not
shown.} 

Results are in Table~\ref{tab:ineff}. I begin with 50 groups since
that size corresponds to much analysis.  CLOGIT is noticeably (20\%)
more accurate for $N=20$   and even when $N=50$ CLOGIT is
still 10\% more accurate for $J=50$; CLOGIT continues to be more
accurate, even though minimally so, even for $N=100$.  In other
words, even though the bias of FELOGIT becomes small (under 5\% when
N reaches 20), there is still  a non-trivial  loss of accuracy in using
FELOGIT instead of CLOGIT. Not surprisingly, these results become more
pronounced when the number of fixed effects estimated is larger
($G=100$) and less pronounced when the number of fixed effects
estimated is smaller ($G=20$). But CLOGIT is always more accurate than
FELOGIT, and substantially more accurate when N is small (say 20 or
under).  

\begin{table}[tb]
\begin{center}
\begin{tabular}{l...}
\cols{1}{l}{N} & \cols{3}{c}{Relative Accuracy$^1$} \\
\cline{2-4} 
\skipc{1} & \cols{1}{c}{$G=20$} & \cols{1}{c}{$G=50$} & \cols{1}{c}{$G=100$} \\
  3 & 2.19  & 2.51 & 3.04  \\
  5 & 1.72  &  1.96 & 2.38 \\
  7 & 1.47 &  1.64 & 2.01 \\ 
 10 & 1.29 &  1.45 & 1.66 \\
 20 & 1.14 &  1.21 & 1.40 \\ 
 30 & 1.09 &   1.17 & 1.27  \\
 50  & 1.06 &  1.10 & 1.15 \\
75 & 1.04 &  1.08  & 1.12 \\
100 & 1.03 &  1.04 & 1.08 \\
 \cline{1-4}
\cols{4}{l}{$^1\frac{\text{RMSE}(\hat{\bb}_{\text{FELOGIT}})}
{\text{RMSE}(\hat{\bb}_{\text{CLOGIT})}}$}
\end{tabular}
\end{center}
\caption{Relative 
   accuracy  of CLOGIT vs LOGIT with fixed
   effects\label{tab:ineff}}
\end{table}

Intuition honed on the continuous dependent variable case would tell
us that only $N$, not $G$, matters. This is because in the continuous
variable case all that matters is the quality of the estimate of the
individual fixed effects, which is only a function of $N$.  But
because of the non-linearity of the logit model, models with more
parameters estimate those parameters less accurately; this is true for
all logit models, not just models with fixed effects. Thus, for
example, when $G=20$ and $N=20$ the relative advantage of CLOGIT over
FELOGIT decreases to 14\%, but for the same $N$, when $G=100$, the
relative accuracy of CLOGIT over FELOGIT increases to 40\%. I stress
that this is not an incidental parameters problem, since the number of
observations is proportional to the number of fixed effects (which is
governed by $N$).

To see how poorly logit handles a huge number of covariates, we can
compare the impact of including many  covariates using both OLS and
LOGIT specifications 
where there is literally no group structure (so no chance of an
incidental parameters problem), but where the number of parameters
estimated is large. Since the linear and logit models are different,
there is no sensible way to compare estimates for the same model. But
it is easy to compare the relative performance of logit and OLS
separately, with the comparison being to a specification with a small
and large number of covariates; this comparison is over a correct specification with a
single covariate and an incorrect one that also includes a large
number of irrelevant 
 covariates. 

Results are shown in  Table~\ref{tab:clreg}. Data were either generated with a
linear or logit specification with literally no group structure.\fn{To keep notation comparable to the previous
  table, results are given in terms of  $N$ and $G$ , but in this data generation
  process all that is relevant is the total number of observations,
  $NG$,  the number of extraneous
  parameters, $G$ and the ratio of observations to parameters
  ($\approx N$).} The data generation process was as simple as
possible. For the logit comparison, a random normal $x$ was generated, this was used to
generate a probability of success using a standard inverse logit
transform ($\gb=1$) and then a binary $y$ was generated as a Bernoulli
random variable.  The continuous $y$ was  simply the latent used
to generate the binary $y$ with a normal error added. The incorrect,
overly large, specification  added $G$ normal variates to the
specifications; these normal variates were generated independently of
$x$ and did not enter the process for generating the latent $y$. 
The table presents the relative RMS error
in estimating \gb in 
 in the logit and linear specifications  with and without the $G$ extraneous variables.\fn{Because of the high variability of the logit
  estimates with some parameter combinations. 
  Table~\ref{tab:clreg} reports  root median square errors. As in previous analyses, data sets were generated 
anew if
  they led to non-convergent logit results. In addition, the logit
  with superfluous variables often dropped observations due to perfect
separation; to keep comparability, the same observations were also
dropped for the univariate logit.}

Clearly the impact of adding many extraneous irrelevant variables is
much more costly for logit than for OLS, with this cost, of course,
becoming less as the ratio of observations to parameters
increases. Note that as $G$ goes from 20 to 100 the number of
observations entering both the logit and OLS estimations grows by a
similar factor of five, so that the ratio of observations to
parameters is constant over the values of $G$. In spite of this, and
in spite of our intuition honed on linear models, logit is much more
sensitive to irrelevant covariates as the number of such covariates
grows, even if the number of observations also grows
proportionally. With 20 irrelevant
variables the accuracy of logit is only really bad until the number of
irrelevant parameters is above  about 5\% of the number of observations;
with 100 irrelevant variables logit is similarly bad when  the number
of irrelevant parameters is over about 1\% of the number of
observations. We usually do not run analyses with so many covariates,
but fixed effects is an exception to this. The problem, to say it
again,  has nothing
to do with grouped data or the incidental parameters problem or fixed
effect \emph{per se}\/ and everything to do with the fact that   non-linear models such as logit are simply more
inaccurate as the number of parameters increases. Obviously this is an
issue with OLS, but to a vastly smaller degree.

\begin{table}[tb]
\centering
\begin{tabular}{r......}
N & \cols{2}{c}{G=20} & \cols{2}{c}{G=50} & \cols{2}{c}{G=100} \\ 
\skipc{1} & \cols{1}{c}{Logit} & \cols{1}{c}{OLS}  & \cols{1}{c}{Logit} & \cols{1}{c}{OLS}  & \cols{1}{c}{Logit} & \cols{1}{c}{OLS} \\
\cmidrule(lr){2-3}  \cmidrule(lr){4-5} \cmidrule(lr){6-7} 
 3  &4.85  &1.26  & 6.77   &1.34   & 9.15 &1.29   \\
 5  &2.03  &1.07  & 2.78   &1.09   & 4.21 &1.08  \\
 7  &1.68  &1.05  & 2.12   &1.11   & 2.87 &1.09  \\
10  & 1.41 & 1.04 &  1.66  & 1.04  &  2.21& 1.08\\
20 &  1.15 & 1.02 &  1.36  & 1.05  &  1.55& 1.02\\
30  & 1.11 & 1.03 &  1.21  & 1.02  &  1.32& 1.01\\
50  & 1.07 & 1.03 &  1.09  & 1.03  &  1.22& 0.99\\
75  & 1.06 & 1.02 &  1.07  & 1.01  &  1.16& 1.03\\
100 & 0.99 & 1.02 &   1.05 &  0.99 &  1.12& 0.99\\
 \cline{1-7}
\end{tabular}
\caption{Comparison of effect of inclusion of $G$ irrelevant covariates
  in logit and OLS with $NG$ observations on relative accuracy of logit and regression
  respectively 
$\frac{\text{RMedianSE}(\hat{\bb}_{\text{G irrelevant covariates}})}
{\text{RMedianSE}(\hat{\bb}_{\text{no irrelevant covariate})}}$\label{tab:clreg}}
\end{table}

With intuition formed on linear models, we often forget how poor
non-linear models estimate parameters when there is either a small
amount of data or a relatively large number of parameters. 
It is interesting that CLOGIT does not have the same problem as
FELOGIT, because CLOGIT conditions  on the actual  number of
successes in a group rather than an estimated group specific
intercept. It is interesting that a method designed to avoid the
incidental parameters problem also has  good finite sample properties
for estimating a logit with fixed effects.

It appears clear that CLOGIT is superior to FELOGIT until 
the number of observations per group  is
quite large, and the superiority of CLOGIT increases monotonically
with the number of groups. CLOGIT is about as fast to estimate as
is FELOGIT, so for estimation of the parameter  of interest, \gb, it
would seem as though there should be a clear preference for
CLOGIT. The only case where this might not be correct is where group
sizes are very large (well into the hundreds), where the CLOGIT model
runs into serious numerical issues. But in such a case FELOGIT will be
fine. Researchers can stick to CLOGIT until their program simply
stops working; at that point, FELOGIT will be fine. Such situations
are rare in published research. 

So why the interest in FELOGIT in this article. This gets back to the
issue of estimating marginal effects. As noted several times, CLOGIT
simply cannot do this. Hence the many researchers interested in sample
marginal effects often resort to the LPMFE, with its attendant
mis-specification issues. The FELOGIT specification deals with the
misspecification. But does the inaccuracy of FELOGIT in practical
situations make FELOGIT a less attractive alternative to LPMFE when a
researcher needs to compute sample average marginal effects? We turn
to this issue in the next section.

\section{FELOT  vs.  LPMFE for estimating marginal
  effects}\label{s:marginal}

To summarize what we have seen so far, CLOGIT is superior to FELOGIT
in general, but CLOGIT does not allow for the estimation of sample
marginal effects. In addition, LPMFE on the entire data set is a
weighted average of zero and the \bb in the NOTALL0 groups. Thus if we
want to compare marginals
 estimated using LPMFE (estimated by OLS) and
LOGITFE estimated by FELOGIT, we should restrict the OLS observations
to the NOTALL0 group (leaving it to analysts whether they then want to
average in 0 for the ALL0 groups). Obviously the LPMFE suffers the defect
that it is not data admissible with a binary dependent variable, but
we have also seen that FELOGIT has poor accuracy properties unless
there are either a very large number of observations per group or a
relatively small number of groups.\fn{The bible for causally oriented
  econometricians, \citen[107]{angrist:bk}, states ``[t]he upshot of
  this discussion is that while a nonlinear model may fit the
  [conditional expectation function] for [limited dependent
  variables] more closely than a linear model, when it comes to
  marginal effects, this probably matters little. This optimistic
  conclusion is not a theorem, but, as in the empirical examples [in
  the book], it seems to be fairly robustly true.'' This section tries
  to add a bit of extra information on this issue in the context of
  fixed effects.}

There is a third estimation strategy available which should improve on
FELOGIT. This consists of first estimating \bbh\ by CLOGIT, and then
running the FELOGIT specification constraining the estimate of  \bbh\
to be that estimated by CLOGIT. This should improve FELOGIT a bit,
since the estimate of  \bbh\ from CLOGIT is more accurate than the
corresponding estimate in the FELOGIT estimation. This procedure,
while it should help, does not solve the problem of FELOGIT estimating
a large number of parameters. 

Simulations to compare the three estimators of sample average marginal
effects were run, generating the data using a logit model, that is, the
best case for FELOGIT over LPMFE; a real world (but unknowable)
comparison would be less favorable to FELOGIT. Data were simulated as
in the previous section, with $N$ and $G$ varied; as noted for other
results, the results reported here did not vary greatly as other
parameters were varied.  Relative accuracy is as in the previous sections, that is the
RMS error of the estimated sample marginal effect around the true
(known from the DGP) marginal effects. For the OLS estimates the
estimated marginal effect is just $\hat{\gb_{\text{OLS}}}$ using the
NOTALL0 groups while for the FELOGIT
estimates it is the average of the $\hat{\gb_{\text{LOGIT}}}\times
\widehat{P(\ygi=1)} \times \widehat{P(\ygi=0)}$ in the same NOTALL0 groups (with the true
marginal effect being the latter term with the known values replacing
the estimated ones); in this section accuracy is always used to mean
accuracy of sample average  marginal effects.

In Table~\ref{tab:cunc} FELOGIT and the constrained FELOGIT are
compared for accuracy. This table shows
that constrained FELOGIT is almost always  more accurate and never
much less accurate than the unconstrained FELOGIT. Thus this article
only compares constrained FELOGIT estimates of marginal effects with their
the LPMFE/OLS  counterparts. The advantage of constraining the FELOGIT
estimator is only relevant for small $N$, so researchers might choose
to estimate the simpler unconstrained estimator when $N$ is, say, 10
or more; this also make estimation of standard errors much
simpler. But for the purposes of this article it is only necessary to
compare the constrained FELOGIT estimator with OLS.

\begin{table}[tb]
\centering
\begin{tabular}{r...}
N & \cols{1}{c}{G=20} & \cols{1}{c}{G=50} & \cols{1}{c}{G=100} \\ 
\cline{2-4} 
 3  &1.54  & 2.00  & 2.45 \\
 5  &1.36  & 1.62  & 1.97 \\
 7  &1.16  & 1.28  & 1.48 \\
10  &1.11  & 1.13  & 1.11\\
20 & 1.00  & 0.99  & 0.91\\
30  & 0.99 &  0.95 & 0.93\\
50  & 0.99 &  0.97 & 0.95\\
75  & 1.00 &  0.98 & 0.94\\
100 & 0.99 &  0.99 & 0.96\\
 \cline{1-4}
\end{tabular}
\caption{Relative accuracy of unconstrained   and constrained FELOGIT for
  estimating marginal effects\\($\frac{\text{RMSE(Sample Average Marginal via
 unconstrained logit}}{\text{RMSE(Sample Average Marginal via constrained
  logit}}$)}
\label{tab:cunc}
\end{table}

Table~\ref{tab:marginals} contains the comparison of the accuracy of
constrained FELOGIT and OLS (stressing that the OLS estimates are only
on the NOTALL0 groups). While there are a few parameter combinations (large $G$, large
$N$ and successes not being rare) where OLS was slightly better than
constrained FELOGIT, the difference in accuracy in these cases was
 less than 5\%. On the other hand,
 constrained FELOGIT is substantially more accurate than OLS when $N$ is small with OLS and FELOGIT providing similar levels of
accuracy by the time $N$ reaches about 20. Constrained FELOGIT's
advantage is also stronger as the number of groups  grows
larger. Hence we can say that constrained FELOGIT is
 essentially always as good or better than OLS for estimating sample
 average marginals (given the DGP studied) with the advantage being
 non-trivial for small $N$ or large $G$. 

\begin{table}[tb]
\centering
\begin{tabular}{r...}
N & \cols{1}{c}{G=20} & \cols{1}{c}{G=50} & \cols{1}{c}{G=100} \\ 
\cline{2-4} 
 3  & 1.30 & 1.73  &       2.10  \\
 5  & 1.19 & 1.44  &       1.73  \\
 7  & 1.08 & 1.19  &       1.36  \\
10  & 1.08 & 1.10      &    1.06 \\
20 &  1.03 & 1.00  &        0.92 \\
30  & 1.01  &0.96  &       0.95 \\
50  & 1.01  &  0.99    &   0.96 \\
 75  & 1.02  & 1.01  &     0.96 \\
100  & 1.02  & 1.00  &      0.98 \\
 \cline{1-4}
\end{tabular}
\caption{Relative accuracy of OLS and constrained FELOGIT for
  estimating marginal effects\\($\frac{\text{RMSE(Sample Average Marginal via
  OLS}}{\text{RMSE(Sample Average Marginal via constrained
  logit}}$). Quartiles of P: .25,.50,.75}
\label{tab:marginals}
\end{table}

Table~\ref{tab:marginals} is a reasonably favorable case for
OLSFE. Because of the imposed DGP, half of the simulated probabilities
lie between 0.25 and 0.75 (symmetrically). Thus, for example the individual
marginal effects differ relatively little; at the median of the data
the true marginal effect is 0.25, while at the top or bottom quartiles
Thus the linear
approximation, which assumes constant marginal effects, is not that
far off. If we change the constant term from zero to minus two, about
half the simulated probability lie between 0.04 and 0.30 (with the
median being .12). At the median the true marginal effect is 0.11,
whereas at the bottom quartile it is 0.04 and at the top quartile it
is 0.21. Table~\ref{tab:marginals1} repeats Table~\ref{tab:marginals}
with the change in constant term.\fn{The lower generated probabilities
  lead to more ALL0 groups which are dropped from both the OLS and
  logit analyses, leading to an effective smaller number of
  groups. Thus in the table the results for $N=3$ are meaningless but
  retained for comparability of tables.} Looking at these simulations,
the results are somewhat less favorable for OLS, particularly  as
the number of groups gets large (at least until the number of
observations per group also gets large). Thus, for example, with 20
observations per group and 100 groups, constrained FELOGIT is about
25\% more efficient than OLS (as compared to OLS being about 8\% more
efficient in the simulations with larger probabilities centered on
.5). While a 25\% gain in efficiency may not appear enormous, this
means that computing marginal effects with OLS is equivalent to
throwing away about one third of your data.

\begin{table}[tb]
\centering
\begin{tabular}{r...}
N & \cols{1}{c}{G=20} & \cols{1}{c}{G=50} & \cols{1}{c}{G=100} \\ 
\cline{2-4} 
 3  & 1.19 & 1.52  &       1.84 \\
 5  & 1.23 & 1.43  &       1.72  \\
 7  & 1.18 & 1.39  &       1.60  \\
10  & 1.15 & 1.32      &    1.46 \\
20 &  1.10 & 1.15  &        1.24 \\
30  & 1.05  & 1.07  &       1.12 \\
50  & 1.05  &  1.04    &   1.04 \\
 75  & 1.04  & 1.01  &     1.04 \\
100  & 1.02  & 1.04  &      1.04 \\
 \cline{1-4}
\end{tabular}
\caption{Relative accuracy of OLS and constrained FELOGIT for
  estimating marginal effects\\($\frac{\text{RMSE(Sample Average Marginal via
  OLS}}{\text{RMSE(Sample Average Marginal via constrained
  logit}}$). Quartiles of P: .04,.12,.30}
\label{tab:marginals1}
\end{table}

One issue that must be borne in mind is that when reporting marginal
effects analysts should also report their uncertainty. This is trivial
in the OLS context. It is only a bit less trivial in the constrained
FELOGIT context, since the second logit in that context assumes the
estimated $\hat{\gb}$ is the true \gb. It is easy correct this using
simulation or resampling methods. Alternatively, since it is
likely that the uncertainty of estimating the fixed effects
substantially dominates the uncertainty of estimating \gb, the FELOGIT
based estimate of the marginals should not be very
anti-conservative. 
\section{Conclusion} \label{s:conc}

The takeaway from this article is fairly simple. Researchers often
require fixed effects specifications to treat unmodeled heterogeneity
which is correlated with the covariates. Such researchers often
either choose CLOGIT or OLS without justification, or present the
results of both. While in many cases both CLOGIT and OLS yield the
same sign and  crossing of the $p<.05$ level, we have seen that
the appropriate comparison for CLOGIT is regression dropping groups
that do not vary on the dependent variable.

We have also seen that much social science data involves fixed effects
where the number of groups is fixed (whatever the number of
observations per group). While the discussion of estimators has been
dominated by the inconsistency of FELOGIT given the incidental
parameters problem, the type of data discussed in this article have  literally
nothing to do with the issues originally raised by
\citen{neyman:econ}. This is not to say that FELOGIT performs well when
the number of effects estimated is large (compared to group size), but
rather the issues have everything to do with the complications of
non-linear estimation and nothing to do with asymptotics in $G$
(which, while perhaps large, is fixed). 

It is the case, however, that CLOGIT yields better estimates of \gb
than does FELOGIT. This is because ClOGIT is conditioning on a known
quantity, the number of successes in a group, whereas FELOGIT suffers
from estimating a large number of extra parameters (as many extra
parameters as there are groups). Thus even though the theoretical
argument for the superiority of CLOGIT in the types of data under
discussion here is not correct, CLOGIT is still the preferred
alternative for estimating such models.

One reason that researchers may prefer OLS to CLOGIT is the former
allows for the computation of sample average marginal
effects. But, while CLOGIT does not allow for such computation,
FELOGIT does allow for such computations. Researchers may not have
considered FELOGIT due to 
a misunderstanding of the incidental parameters problem.

But we have seen that FELOGIT can be improved by constraining the
estimate of \gb to the superior estimate which is yielded by
CLOGIT. This generally improves the accuracy, often non-trivially, of estimating marginal
effects (and almost never hurts). Thus researchers can use this
procedure if the estimation of sample marginal effects is
required. Such a procedure is superior to OLS when the number of
observations per group is small or the number of groups is large; in
addition, there is little cost to always using constrained CLOGIT
 over OLS.

Does this advice generalize to other ways of simulating data? Of
course this is impossible to know in general. Playing with a variety
of such processes, it has been hard to find any where OLS is superior
to a version of logit. Has it been shown that a variant of logit is always
\emph{much}\/ better for estimating marginal effects? No. But in some
situations (where probabilities of success are far from .5) logit is
non-trivially better. Thus the final
takeaway is that researchers should use a variant of FELOGIT for
estimating marginal effects unless the complications of so doing (say
for issues related to endogeneity) are great. This is a slight to
moderate 
amendment to the advice due to \citeauthor{angrist:bk} cited earlier.
\clearpage
\bibliographystyle{pa}
\bibliography{neal}

\clearpage

\end{document}